\documentclass[floats,showpacs,pra,twocolumn,superscriptaddress]{revtex4}%
\usepackage{amssymb}
\usepackage{amsmath}
\usepackage{epsfig}
\usepackage{graphicx}
\usepackage{amsfonts}%
\providecommand{\U}[1]{\protect\rule{.1in}{.1in}}
\begin{document}
\title{Survival Probability of a Local Excitation in a Non-Markovian Environment:
Survival Collapse, Zeno and Anti-Zeno effects.}
\author{E. Rufeil Fiori}
\email{rufeil@famaf.unc.edu.ar}
\affiliation{Facultad de Matem\'{a}tica, Astronom{\'{\i}}a y F{\'{\i}}sica, and Instituto
de F\'{\i}sica (CONICET), Universidad Nacional de C\'{o}rdoba, Ciudad
Universitaria, 5000, C\'{o}rdoba, Argentina.}
\author{H. M. Pastawski}
\email{horacio@famaf.unc.edu.ar}
\affiliation{Facultad de Matem\'{a}tica, Astronom{\'{\i}}a y F{\'{\i}}sica, and Instituto
de F\'{\i}sica (CONICET), Universidad Nacional de C\'{o}rdoba, Ciudad
Universitaria, 5000, C\'{o}rdoba, Argentina.}
\begin{abstract}
The decay dynamics of a local excitation interacting with a non-Markovian
environment, modeled by a semi-infinite tight-binding chain, is exactly
evaluated. We identify distinctive regimes for the dynamics. Sequentially: (i)
early quadratic decay of the initial-state survival probability, up to a
spreading time $t_{S}$, (ii) exponential decay described by a self-consistent
Fermi Golden Rule, and (iii) asymptotic behavior governed by quantum diffusion
through the return processes and leading to an inverse power law decay. At
this last cross-over time $t_{R}$ a survival collapse becomes possible. This
could reduce the survival probability by several orders of magnitude. The
cross-overs times $t_{S}$ and $t_{R}$ allow to assess the range of
applicability of theFermi Golden Rule and give the conditions for the
observation of the Zeno and Anti-Zeno effect.

\end{abstract}
\startpage{1}
\maketitle

\section{Introduction}

The decay of an unstable local state is usually described, within a Markovian
approximation, by an exponential decay with a rate given by the Fermi Golden
Rule (FGR). However, this description contains approximations that leave aside
some intrinsically quantum behaviors \cite{Raizen}. Indeed, works on models
for nuclei, composite particles \cite{Kal} and excited atoms \cite{atoms},
predict that the exponential decay does not hold for very short and very long
times, and this exponential decay may shows superimposed beats.

In Ref. \cite{RP06} we presented an exactly-solvable model describing the
evolution of a surface excitation in a semi-infinite chain. Physical
realizations of one-dimensional systems are provided by electron transport in
superlattices \cite{GC07}, discrete diffraction in photonic crystals
\cite{Lon06}, and spin excitations in a chain of nuclear spins under an XY
interaction \cite{MBS+97} or under double quantum interaction \cite{HAP}.
Here, we quantify and interpret the short and long time limits, $t_{S}$ and
$t_{R}$, of the FGR. We identify three time regimes for the survival
probability $P_{00}(t)$. Initially the decay is quadratic and it holds up to a
time $t_{S}$. From $t_{S}$ to $t_{R}$ it is exponential, and finally, for long
times, it follows a power law. The time $t_{S}$ gives an upper bound to the
time interval at which repetitive projection measurements could lead to a
Quantum Zeno Effect (\cite{MS77}, \cite{FNP01}). On the other hand, at $t_{R}%
$, a dip in $P_{00}(t)$ of several orders of magnitude may occur. This
\textit{survival collapse} is identified with a destructive interference
between the \textit{pure survival} amplitude, i.e., an exponential decay
amplitude, and a \textit{return} amplitude, which is usually neglected because
it arises from memory effects in the environment. This destructive
interference can be used to obtain an anti-Zeno effect \cite{FGR01}, where the
decay rate is strongly enhanced by repeated projective measurements with
period $t_{R}$.

\section{Survival probability}

\begin{figure}
[h]
\begin{center}
\includegraphics[
trim=0.714134in 0.275707in 0.178627in 0.463779in,
height=2.7527in,
width=3.3183in
]%
{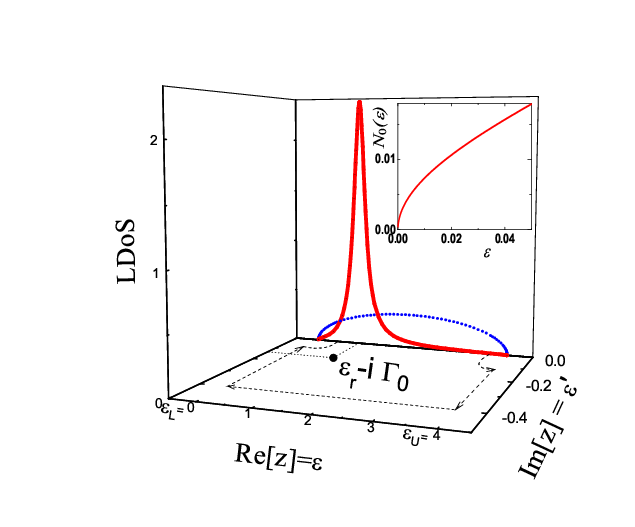}%
\caption{(color online) Local Densities of States (LDoS) in the complex plane
$z=\varepsilon+\mathrm{i}\varepsilon^{\prime}$. $\varepsilon_{L}$ and
$\varepsilon_{U}$ are the lower and upper band-edges, respectively. The solid
line is $N_{0}\left(  \varepsilon\right)  $ for a semi-infinite chain with a
surface impurity. The dotted line is $N_{1}^{(0)}\left(  \varepsilon\right)  $
for a semi-infinite homogeneous chain, and in the inset is shown the lower
band-edge of $N_{0}\left(  \varepsilon\right)  $. The pole appears in
$\varepsilon_{r}-\mathrm{i}\Gamma_{0}$. The integration path is shown with
dashed lines; it consist of four straight lines and two arcs (which avoid the
band-edges singularities).}%
\label{Fig_No}%
\end{center}
\end{figure}

The evolution of a state $\left\vert 0\right\rangle $ weakly coupled to a set
of states which defines the \textquotedblleft environment\textquotedblright,
is described by the survival probability%
\begin{align}
P_{00}\left(  t\right)   &  =\left\vert \left\langle 0\right\vert
\exp[-\mathrm{i}\mathcal{H}t/\hbar]\left\vert 0\right\rangle \theta\left(
t\right)  \right\vert ^{2}\\
&  \equiv\hbar^{2}\left\vert G_{00}^{R}\left(  t\right)  \right\vert ^{2},\\
&  =\hbar^{2}\left\vert \int\frac{\mathrm{d}\varepsilon}{2\pi\hbar}G_{00}%
^{R}(\varepsilon)\exp[-\mathrm{i}\varepsilon t/\hbar]\right\vert ^{2},\\
&  =\left\vert \theta\left(  t\right)  \int_{-\infty}^{\infty}\mathrm{d}%
\varepsilon\text{ }N_{0}\left(  \varepsilon\right)  \exp[-\mathrm{i}%
\varepsilon t/\hbar]\right\vert ^{2}, \label{Eq_PdeN}%
\end{align}
where $G_{00}^{R}\left(  t\right)  $ is the retarded single particle Green's
function and $N_{0}\left(  \varepsilon\right)  $ is the Local Density of
States (LDoS). This is evaluated expanding the initial condition in the
eigenstates $\left\vert k\right\rangle $ of the Hamiltonian $\mathcal{H}$, or
by using the energy representation of the Green's function $G_{00}^{R}\left(
\varepsilon\right)  $,%
\begin{align}
N_{0}\left(  \varepsilon\right)   &  \equiv\sum\nolimits_{k}\left\vert
\left\langle 0|k\right\rangle \right\vert ^{2}\delta\left(  \varepsilon
-\varepsilon_{k}\right)  ,\\
&  =-1/\pi\operatorname{Im}G_{00}^{R}(\varepsilon).
\end{align}
If the spectrum is bounded, Eq. (\ref{Eq_PdeN}) can be calculated using the
residue theorem with the path shown in Fig. \ref{Fig_No}.%

Resonances appear as poles of the analytical continuation $N_{0}(z)\equiv
N_{0}(\varepsilon+\mathrm{i}\varepsilon^{\prime})$ in the lower complex
semi-plane. A \textit{well defined resonance }appears when an initially
unperturbed state of energy $\varepsilon_{0}=\left\langle 0|\mathcal{H}%
|0\right\rangle $, far enough from the band-edge, is weakly coupled to a
continuum, i.e., the expansion of $\left\vert 0\right\rangle $ in terms of the
eigenstates has a small breath $\Gamma_{0}$ around an energy $\varepsilon
_{r}=\varepsilon_{0}+\Delta_{0}$, where $\Delta_{0}=\Delta\left(
\varepsilon=\varepsilon_{r}\right)  $ is a small shift due to the interaction.
This condition excludes out-of-band resonances, virtual states and localized
eigenstates \cite{DBP08}. Then,%
\begin{align}
P_{00}(t)  &  =|\underset{\mathrm{SC-FGR}}{\underbrace{~a~~\mathrm{e}%
^{-(\Gamma_{0}+\mathrm{i}\varepsilon_{r})t/\hbar}}}\nonumber\\
&  +\int\limits_{0}^{\infty}\mathrm{e}^{-\varepsilon^{\prime}t/\hbar}\left[
\mathrm{e}^{-\mathrm{i}\text{ }\varepsilon_{L}t/\hbar}N_{0}(\varepsilon
_{L}-\mathrm{i}\varepsilon^{\prime})\right. \nonumber\\
&  \underset{\text{return correction from quantum diffusion}}{\underbrace
{\left.  -\mathrm{e}^{-\mathrm{i}\text{ }\varepsilon_{U}t/\hbar}%
N_{0}(\varepsilon_{U}-\mathrm{i}\varepsilon^{\prime})\right]  \text{\textrm{d}%
}\varepsilon^{\prime}}}|^{2}, \label{Eq_Poo}%
\end{align}
where $a=2\pi\mathrm{i}$ $\lim_{z\rightarrow\varepsilon_{r}-\mathrm{i}%
\Gamma_{0}}$ $\left[  (z-\varepsilon_{r}+\mathrm{i}\Gamma_{0})\text{ }%
N_{0}(z)\right]  $ is the pole residue. $P_{00}$ presents two separate
contributions for the decay. The first term (the pole contribution) of
Eq.(\ref{Eq_Poo}) supersedes the usual FGR approximation since it has a
pre-exponential factor ($A\equiv|a|^{2}\gtrsim1$) and an exact rate of decay
$\Gamma_{0}$, i.e., this result is a \textit{self-consistent Fermi Golden
Rule} (SC-FGR). This term is the dominant one for a wide range of times,
leading to%
\begin{equation}
P_{00}\left(  t\right)  \approx A\exp\left(  -2\Gamma_{0}t/\hbar\right)
\label{Eq_Poo_exp}%
\end{equation}
By analogy with the self-diffusion process in a classical Markov chain, the
exponential in Eq. (\ref{Eq_Poo}) is identified with a \textit{pure
survival\ }amplitude\textit{. }Within the same analogy\textit{,} the second
term (the integration path contribution) will be called\textit{\ return\ }%
amplitude. The \textit{\textquotedblleft quantum diffusion}\textquotedblright%
\ described by this term dominates for long times and brings out the details
of the spectral structure of the environment.

\subsection{Short time regime}

\begin{figure}
[h]
\begin{center}
\includegraphics[
height=1.8879in,
width=2.0366in
]%
{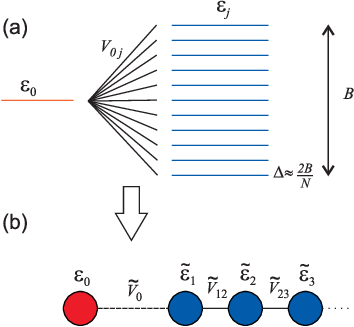}%
\caption{(color online) (a) Single state of energy $\varepsilon_{0}$ coupled
by$\ V_{0j}$ to $\mathrm{N}$ states of energy$\ \varepsilon_{j}$, spread over
a bandwidth $B$\ with mean-level space of $\Delta\approx2B/N$,\ that defines
the environment. (b) Equivalent semi-infinite chain with $\varepsilon
_{0},\widetilde{V_{0}},\widetilde{\varepsilon_{1}},$ defined in text.}%
\label{Fig_niveles}%
\end{center}
\end{figure}

The second term of Eq. (\ref{Eq_Poo}) is fundamental for the normalization at
very short times. Both terms combine to provide the initial quadratic decay
required by the perturbation theory%
\begin{equation}
P_{00}\left(  t\right)  =1-\left\langle (\varepsilon-\varepsilon_{r}%
)^{2}\right\rangle _{N_{0}}t^{2}/\hbar^{2}+\cdots,
\end{equation}
where $\left\langle (\varepsilon-\varepsilon_{r})^{2}\right\rangle _{N_{0}}$
is the second moment of the LDoS $N_{0}\left(  \varepsilon\right)  $. This
expansion holds up to the characteristic time $t_{S}$. Let us consider a
single state of energy $\varepsilon_{0}$\ coupled by $V_{0,j}$ to an
\textit{environment} defined by $\mathrm{N}$ states of energy $\varepsilon
_{j}$ spread over a bandwidth $B$, as shows Fig.(\ref{Fig_niveles}-a).%

A semi-infinite chain can be obtained from this system by using the recursion
method \cite{HHK72}, a variant of Lanczos tridiagonalization scheme, as is
shown in Fig.(\ref{Fig_niveles}-b). The first two states are:%
\begin{align}
\widetilde{\varepsilon_{0}}  &  =\varepsilon_{0};\text{ \ \ }\widetilde
{\left\vert 0\right\rangle }=\left\vert 0\right\rangle ,\\
\widetilde{\varepsilon_{1}}  &  =\frac{\sum\nolimits_{j=1}^{N}V_{j,0}%
\ \varepsilon_{j}}{\widetilde{V_{0}}^{2}};\text{ }\widetilde{\left\vert
1\right\rangle }=\frac{\sum\nolimits_{j=1}^{N}V_{0\text{,}j}\left\vert
j\right\rangle }{\widetilde{V_{0}}},\\
\widetilde{V_{0}}  &  =\sqrt{\sum\nolimits_{j}^{N}\left\vert V_{0,j}%
\right\vert ^{2}}.
\end{align}
Here, the local second moment of the Hamiltonian is $\widetilde{V_{0}}^{2}$,
leading to%
\begin{equation}
P_{00}\left(  t\right)  =1-\widetilde{V_{0}}^{2}t^{2}/\hbar^{2}+\cdots
\label{Eq_Poo_short}%
\end{equation}
There is a simple expression extrapolating Eqs. (\ref{Eq_Poo_short}) and
(\ref{Eq_Poo_exp}), \cite{FI01}:%
\begin{equation}
P_{00}\left(  t\right)  \approx\exp[\left(  1-\sqrt{1+(t/t_{S})^{2}}\right)
2\Gamma_{0}t_{S}/\hbar],
\end{equation}
with,%
\begin{equation}
t_{S}=\hbar\frac{\Gamma_{0}}{\widetilde{V_{0}}^{2}}. \label{Eq_ts}%
\end{equation}
This yields Eq. (\ref{Eq_Poo_short}) for $t\ll t_{S}$ at the lowest order. In
contrast, for $t\gg t_{S}$ it yields the SC-FGR of Eq. (\ref{Eq_Poo_exp}),
with $A\approx\exp(2\Gamma_{0}^{2}/\widetilde{V}_{0}^{2})$\ valid for
$|\widetilde{V}_{0}|\ll B$. Therefore, as was remarked by Pascazio et al.
\cite{FNP01}, the upper limit for the quadratic behavior is not $\hbar
/\widetilde{V}_{0}$, as one might expect,\ but rather the much shorter time
$t_{S}$. A useful interpretation of $t_{S}$ can be drawn from the Green's
function \cite{PM01}:%
\begin{equation}
G_{00}^{R}(\varepsilon)=\frac{1}{\varepsilon-\varepsilon_{0}-\widetilde{V}%
_{0}^{2}G_{\widetilde{1}\widetilde{1}}^{R\left(  0\right)  }(\varepsilon)},
\label{Eq_Goo2d}%
\end{equation}
where $G_{\widetilde{1}\widetilde{1}}^{R\left(  0\right)  }(\varepsilon)$
corresponds to a semi-infinite chain in absence of 0th-site. Taking
$G_{\widetilde{1}\widetilde{1}}^{R\left(  0\right)  }\left(  \varepsilon
=\varepsilon_{0}\right)  $ Eq.(\ref{Eq_Goo2d}) gives the FGR: $\Gamma
_{FGR}\approx\pi\widetilde{V}_{0}^{2}N_{\widetilde{1}}^{\left(  0\right)
}\left(  \varepsilon_{0}\right)  $. Replacing it in Eq. (\ref{Eq_ts}), we get%
\begin{equation}
t_{S}\approx\hbar\pi N_{\widetilde{1}}^{\left(  0\right)  }\left(
\varepsilon_{0}\right)  , \label{Eq_ts_N}%
\end{equation}
only determined by $N_{\widetilde{1}}^{\left(  0\right)  }\left(
\varepsilon_{0}\right)  $,\ the LDoS at the 1st-site of the\ unperturbed
environment, evaluated at $\varepsilon_{0}$. In turns, $\hbar N_{\widetilde
{1}}^{\left(  0\right)  }\left(  \varepsilon_{0}\right)  $ represents the time
scale \cite{PM01} at which an excitation built from the decay, decays into
the\ rest of the environment. Therefore, the return to the 0th-site, required
to build up the quadratic decay, becomes less appreciable than the escape
towards the chain, leading to the fast exponential decay of the survival probability.

\subsection{Long time regime: Survival collapse}

For long times, only small values of $\varepsilon^{\prime}$ contribute to the
integral of the second term in Eq.(\ref{Eq_Poo}). This restricts the
integration to a range near the band-edges. Then, taking into account
Eq.(\ref{Eq_PdeN}) and performing the Fourier transform retaining only the Van
Hove singularities at these edges, we get the power law decay at long times.
This second term dominates $P_{00}\left(  t\right)  $ because its decay is
slower than the exponential one. The relative participation on the LDoS at
each edge is $\beta=[(\varepsilon_{r}-\varepsilon_{L})^{2}+\Gamma_{0}%
^{2}]/[(\varepsilon_{U}-\varepsilon_{r})^{2}+\Gamma_{0}^{2}]$. Collecting both
edge contributions gives%
\begin{align}
P_{00}(t)  &  \approx\left[  1+\beta^{2}-2\beta\cos(Bt/\hbar)\right]
\nonumber\\
&  \times\left\vert \int\text{\textrm{d}}\varepsilon^{\prime}\mathrm{e}%
^{-\varepsilon^{\prime}t/\hbar}N_{0}(\varepsilon_{L}-\mathrm{i}\varepsilon
^{\prime})\right\vert ^{2}. \label{Eq_Plargo}%
\end{align}
This means that the long time behavior is just the power law multiplied by a
factor containing a modulation with frequency $B/\hbar$. Eq. (\ref{Eq_Poo})
shows that the survival amplitude of the local excitation recognizes two
alternative pathways: the \textit{pure\ survival} (pole contribution), and the
returning pathways where the excitation has decayed and explored the
environment. Then, there is an interference term that becomes important when
both amplitudes are of the same order. It is precisely at this cross-over time
$t_{R}$ between the exponential regime and the power law regime when the
interference term can produce\ a \textit{survival collapse}, i.e.,
$P_{00}\left(  t\right)  $ nearly cancels out. This effect is seen as a
pronounced dip in Fig. \ref{Fig_Poo}.

We also note that if the unperturbed energy state $\varepsilon_{0}$ is exactly
at the center of the band, $\beta=1$, the pure return probability presents
periodicals zeros barely compensated by the small pure survival probability.
This should not be confused with the survival collapse discussed above, which
may yield an exact zero in $P_{00}\left(  t\right)  $.

\section{Semi-infinite chain: exact solution}

\begin{figure}
[tbh]
\begin{center}
\includegraphics[
trim=0.279805in 0.151753in 0.219281in 0.228485in,
height=2.3246in,
width=3.0355in
]%
{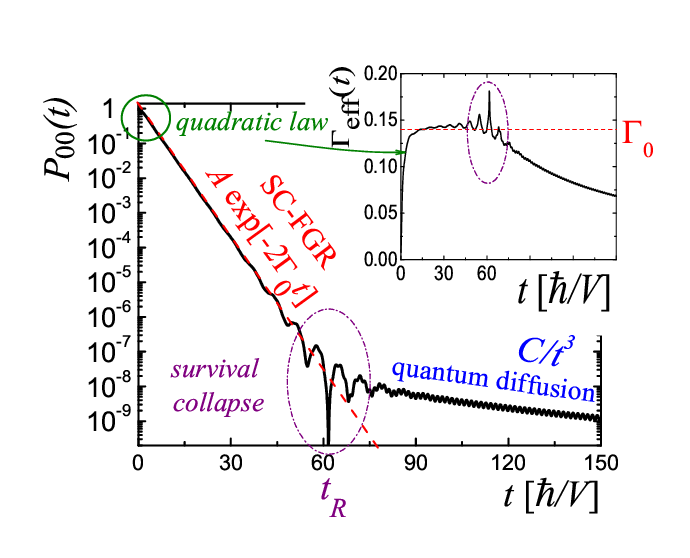}%
\caption{(color online) Survival probability for a semi-infinite chain with
$\varepsilon_{0}/V=1$, $V_{0}/V=0.4$, that leads to a resonance in
$\varepsilon_{r}/V=0.9$, $\Gamma_{0}/V=0.14$. The inset shows $\Gamma
_{eff}\left(  t\right)  $ (solid line) and $\Gamma_{0}$ (dashed line).}%
\label{Fig_Poo}%
\end{center}
\end{figure}

Let us focus on a tight-binding Hamiltonian shown in Fig. (\ref{Fig_niveles}%
-b) with hoppings $\widetilde{V}_{0}=V_{0}$, $\widetilde{V}_{j,j+1}=V$ and
site energies $\varepsilon_{0}$ and $\varepsilon_{j}=2V$ for $j>0$:%
\begin{align}
\mathcal{H}  &  =\left\vert 0\right\rangle \varepsilon_{0}\left\langle
0\right\vert -\left(  \left\vert 0\right\rangle V_{0}\left\langle 1\right\vert
+c.c.\right) \nonumber\\
&  +%
{\displaystyle\sum\limits_{n}}
\left(  \left\vert n\right\rangle 2V\left\langle n\right\vert -\left\vert
n\right\rangle V\left\langle n+1\right\vert +c.c.\right)  . \label{Eq_TBH}%
\end{align}
This defines a continuous spectrum $[\varepsilon_{L}=0,\varepsilon_{U}=4V=B]$
and a well defined resonance for $V_{0}\ll V$. We first summarize the results
in Ref. \cite{RP06}. The LDoS factorizes as a pure Lorentzian around
$\varepsilon_{r}\pm\mathrm{i}\Gamma_{0}$, and $N_{1}^{(0)}(\varepsilon)$:%
\begin{equation}
N_{0}\left(  \varepsilon\right)  =\frac{V^{2}}{\Gamma_{c}}\frac{\Gamma_{0}%
}{\left(  \varepsilon_{r}-\varepsilon\right)  ^{2}+\Gamma_{0}^{2}}N_{1}%
^{(0)}(\varepsilon),
\end{equation}
with%
\begin{align}
N_{1}^{\left(  0\right)  }(\varepsilon)  &  =\frac{16\Gamma\left(
\varepsilon\right)  }{\pi B^{2}}\theta\left(  |\varepsilon-2V|\right)  ;\text{
}\Gamma\left(  \varepsilon\right)  =\frac{\sqrt{\varepsilon}\sqrt
{B-\varepsilon}}{2},\\
\varepsilon_{r}  &  =\varepsilon_{0}+\Delta_{0};\text{ }\Delta_{0}=\frac
{V_{0}^{2}}{V^{2}-V_{0}^{2}}\frac{\varepsilon_{0}-2V}{2},\\
\Gamma_{0}  &  =\frac{V_{0}^{2}}{V^{2}-V_{0}^{2}}\Gamma_{c};\text{ }\Gamma
_{c}=\sqrt{V^{2}-V_{0}^{2}-\left(  \frac{\varepsilon_{0}-2V}{2}\right)  ^{2}}.
\label{Eq_Gamma0.}%
\end{align}
The solution of Eq.(\ref{Eq_PdeN}) results in:%
\begin{equation}
P_{00}(t)\approx\left\{
\begin{array}
[c]{c}%
1-\left(  V_{0}t/\hbar\right)  ^{2},\text{ }t<t_{S}\\
A\exp(-2\Gamma_{0}t/\hbar),\text{ }t_{S}<t<t_{R}\\
C[1-\frac{2\beta}{1+\beta^{2}}\sin\left(  Bt/\hbar\right)  ][\hbar
/(\Gamma(\varepsilon_{r})t)]^{3},t_{R}<t
\end{array}
\right.  \label{Eq_Psemi}%
\end{equation}
with $\beta$\ as defined in section 2 and%
\begin{align}
A  &  =\frac{\sqrt{\varepsilon_{r}^{2}+\Gamma_{0}^{2}}\sqrt{\left(
B-\varepsilon_{r}\right)  ^{2}+\Gamma_{0}^{2}}}{4\Gamma_{c}^{2}},\\
C  &  =\frac{V_{0}^{4}V\Gamma(\varepsilon_{r})^{3}\left(  1+\beta^{2}\right)
}{4\pi\left(  V^{2}-V_{0}^{2}\right)  ^{2}\left(  \Gamma_{0}^{2}%
+\varepsilon_{r}^{2}\right)  ^{2}}.
\end{align}
Notice that the cubic power law decay at long times follows from the
$\sqrt{\varepsilon}$ dependence of the LDoS near the band edge (see inset of
Fig. (\ref{Fig_No})). Fig. (\ref{Fig_Poo}) shows Eq. (\ref{Eq_Psemi}) for
$V_{0}/V=0.4$ and $\varepsilon_{0}/V=1$. One obtain an alternative
representation of Eq.(\ref{Eq_Poo}) by introducing an effective decay rate
$\Gamma_{\mathrm{eff}}\left(  t\right)  =-\hbar/(2t)\ln P_{00}(t)$
\cite{FNP01} whose deviation from $\Gamma_{0}$ is a signature of
non-exponential decay. This is shown in the inset of Fig. (\ref{Fig_Poo}).
There, the survival collapse is a pronounced peak in $\Gamma_{\mathrm{eff}%
}\left(  t\right)  $.%

In order to obtain the characteristic time $t_{R}$ in the weak coupling limit
we solve iteratively the equality between the exponential and the power law
decay (averaged in a period), starting with $\hbar/2\Gamma_{0}$. Since, for
$\varepsilon_{0}$ close to the center of the band $\sqrt{A/C}\approx
\sqrt{32\pi}V/\Gamma_{0}$ and $\Gamma\left(  \varepsilon_{r}\right)  \approx
V$, we obtain%
\begin{equation}
t_{R}^{(0)}=a_{1}\frac{\hbar}{\Gamma_{0}}\ln\left(  a_{2}\frac{B}{4\Gamma_{0}%
}\right)  , \label{Eq_tR}%
\end{equation}
where $a_{1},a_{2}\gtrsim1$ are constants that depend on the Van Hove
singularity $N_{0}\left(  \varepsilon\right)  \sim\left(  \varepsilon
-\varepsilon_{L}\right)  ^{\nu}$ and other details of the model. For a
semi-infinite chain $a_{1}=\nu+2=5/2$ and$\ a_{2}=\sqrt[5]{4\pi}\sim1.6$. By
choosing the parameters $V_{0},\varepsilon_{0}$ as above, this characteristic
time results in $t_{R}^{(0)}\approx41$ $[\hbar/V]$, which is somewhat smaller
than the exact time $t_{R}\approx62$ $[\hbar/V]$. Just the next order of
iteration gives a much better approximation $t_{R}^{(1)}\approx67$ $[\hbar
/V]$. Also, by using these parameters, Eq. (\ref{Eq_ts_N}) results in
$t_{S}\approx0.8$ $[\hbar/V]$ which is a good bound for the short time scale.

In the range of quadratic decay, recursive projective measurement of state
$\left\vert 0\right\rangle $ at a time interval $\tau_{\phi}$ would produce a
\textit{deceleration} of the decay, i.e., a Quantum Zeno effect (see, for
example, Fig. 1 in \cite{PU98}). Our results above provides a convenient upper
bound, $\tau_{\phi}<t_{S}$, for this time scale.

\begin{figure}
[tbh]
\begin{center}
\includegraphics[
trim=0.312020in 0.091047in 0.052566in 0.200601in,
height=2.3826in,
width=2.9603in
]%
{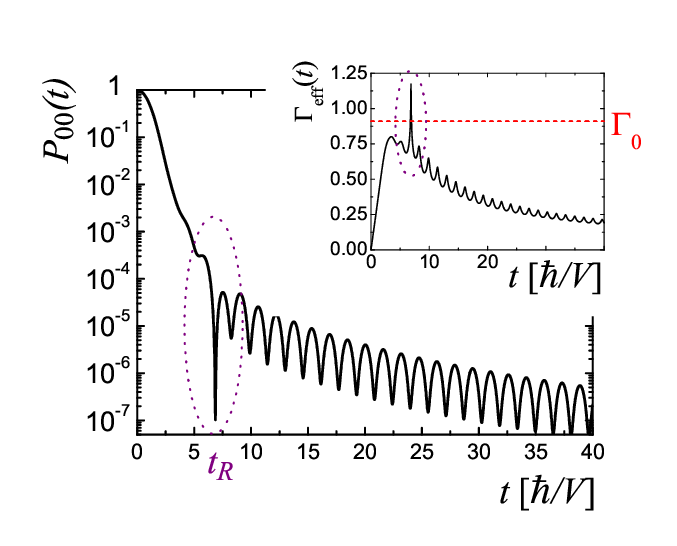}%
\caption{(color online) Survival probability for a semi-infinite chain with
$\varepsilon_{0}/V=1.8$, $V_{0}/V=0.77$. The inset shows $\Gamma_{eff}\left(
t\right)  $ (solid line) and $\Gamma_{0}$ (dashed line).}%
\label{Fig_Poo3}%
\end{center}
\end{figure}

The survival collapse can also occur at the strong coupling limit at the
cross-over between the short time regime and the power law decay. Fig.
(\ref{Fig_Poo3}) shows $P_{00}\left(  t\right)  $ and $\Gamma_{\mathrm{eff}%
}\left(  t\right)  $ for $\varepsilon_{0}/V=1.8$ and $V_{0}/V=0.77$, which
yields to $t_{R}\approx$ $6.8\ [\hbar/V]$. In this case, recursive projective
measurement at a time interval $\tau_{\phi}\approx t_{R}$ can make the
survival probability much smaller. Then the survival collapse enables an
\textit{acceleration} of the decay induced by repetitive observations, i.e.,
an anti-Zeno effect \cite{FNP01}.%

\section{Conclusions}

We studied the dynamics of a local excitation in a system in which full memory
effects at the environment are included. We obtain the time limits where the
non-exponential behavior of the survival probability shows up. The evolution
starts with the expected quadratic decay, which holds up to a time $t_{S}%
$\ (Eq. (\ref{Eq_ts_N})) determined by the density of the first state of the
environment in absence of the initial state.\emph{\ }This time gives an upper
bound to the interval at which repetitive projection measurements leads to a
Quantum Zeno effect. In the weak coupling limit the decay follows the usual
FGR exponential, but with a corrected rate and a pre-exponential factor, i.e.,
the SC-FGR. At long times we get a power law decay controlled by non-Markovian
return processes. We also visualized a survival collapse at time $t_{R}$ (Eq.
(\ref{Eq_tR})) as a destructive interference between the pure survival
amplitude and the return amplitude. This last arises from pathways that
explore the environment before returning. Given that a survival collapse
occurs, one can use repetitive\ projective measurements with a period
$\tau_{\phi}\approx t_{R}$ to achieve an anti-Zeno effect.

\bigskip

The authors acknowledge financial support from CONICET, SeCyT-UNC and ANPCyT
as well as hospitality of Abdus-Salam ICTP.

\end{document}